\documentclass[twocolumn,pra,aps,amssymb,groupaddress,showpacs]{revtex4}
\usepackage{amsmath,amssymb,amsfonts}
\usepackage[usenames]{color}
\usepackage{graphicx,tikz,pgf,epsf}

\newcommand{\qed}{\hfill \rule{2mm}{2mm}}
\newcommand{\pf}{{\bf Proof: }}
\newtheorem{fact}{Fact}
\newtheorem{theorem}{Theorem}
\newtheorem{lemma}{Lemma}
\newtheorem{proposition}{Proposition}

\newtheorem{example}{Example}

\begin{document}
\title{Efficient quantum algorithm to construct arbitrary Dicke states}
\author{Kaushik Chakraborty$^1$, Byung-Soo Choi$^2$, Arpita Maitra$^1$ and Subhamoy Maitra$^1$}
\affiliation{$^1$Applied Statistics Unit, Indian Statistical Institute, 203 B T Road, Kolkata 700 108, India,\\
email: kaushik.chakraborty9@gmail.com, arpita76b@rediffmail.com, subho@isical.ac.in\\
$^2$Department of Electrical and Computer Engineering, Duke University, Durham, NC 27708, USA, email: bschoi3@gmail.com}
\begin{abstract}
In this paper, we study efficient algorithms towards the construction of 
any arbitrary Dicke state. Our contribution is to use proper symmetric Boolean 
functions that involve manipulations with Krawtchouk polynomials. 
Deutsch-Jozsa algorithm, Grover algorithm and the parity measurement technique 
are stitched together to devise the complete algorithm. Further, motivated by 
the work of Childs et al (2002), we explore how one can plug the biased 
Hadamard transformation in our strategy. Our work compares fairly with the 
results of Childs et al (2002).
\end{abstract}
\maketitle

\section{Introduction}
Multipartite entanglement is one of the important areas in the field of quantum
information that has many applications including quantum secret sharing. 
In this paper, we focus on the Dicke states~\cite{PhysRev.93.99}, which are 
useful 
building blocks in realizing multipartite entanglement. The $n$-qubit weight 
$w$ Dicke state, $|D^n_w\rangle$, is the equal superposition of all $n$-qubit 
states of weight $w$. We refer to~\cite{dickold,d2,d3,phyb,d1,d4,d5} and the 
references therein for detailed discussion. 

After the invention of quantum information, many experimental setups have been
proposed and tested to verify some theoretical properties. Most of experiments
have been focused on the test of multipartite entanglement such as EPR, GHZ,
and W states. Since the result of experimental tests depends on the steps for
preparing, processing, and measuring, all steps should be refined as much as
possible. Among them, the first priority is to prepare the target state with
very high fidelity and with efficiency. In this work, therefore, we also focus
on the efficient way to prepare certain multipartite quantum state.

In line of GHZ and W states, we have the 
Dicke state, $|D^n_w\rangle$, which an equal superposition state of all
$n$-qubit states of weight $w$. Actually, Dicke state is more general state
than GHZ and W states since W state is $|D^n_1\rangle$ and GHZ state is the
superposition of $|D^n_0\rangle$ and $|D^n_n\rangle$. Therefore, the
preparation method for Dicke state can be utilized for other general case as
well. At the same time, similar to the above reason, Dicke state can be
utilized for many applications such as secret sharing \cite{d5} and quantum
networking \cite{dd1}.
Related to this, some previous works have been done that focussed on the
experimental ways to prepare six-qubit Dicke state \cite{d5,d4} with fidelity
$0.654\pm0.024$ and $0.56\pm0.02$, respectively.

While the main focus from the viewpoint of experimental physics is 
to actually provide the implementation of specific Dicke states, our focus is 
from theoretical algorithmic angle and the only result presented 
in this direction appeared in~\cite{dickold}. In this work, we show how one 
can efficiently construct Dicke states by using the combinatorial properties 
of symmetric Boolean functions, two well-known quantum algorithms, and the 
generalized parity measurement. By efficient, we mean that the resource 
requirements in terms of quantum circuits and number of execution steps is 
poly$(n)$ to obtain $|D^n_w\rangle$.

Let us consider $n$-qubit states in the computational basis $\{0, 1\}^n$ that 
can be written in the form $\sum_{x \in \{0, 1\}^n} a_x |x\rangle$, 
where $\sum_{x \in \{0, 1\}^n} |a_x|^2 = 1$. Thus, $x$ can also be interpreted 
as a binary string and the number of $1$'s in the string is called the 
(Hamming) weight of $x$ and denoted as $wt(x)$. Based on this an arbitrary 
Dicke state can be expressed as follows:
$$|D^n_w\rangle =\sum\limits_{x\in\{0,1\}^n,wt(x)=w}\frac{1}{\sqrt{\binom{n}{w}}}|x\rangle.$$
Let us also define a symmetric $n$-qubit state as $$|S^n\rangle=\sum_{x\in\{0,1\}^n}a_{wt(x)}|x\rangle, \mbox{ where } \sum_{i=0}^n\binom{n}{i}|a_i|^2=1.$$

First, we show how one can prepare a symmetric $n$-qubit state with the 
property that $\binom{n}{w}|a_w|^2$ is $\Omega(\frac{1}{\sqrt{n}})$ by using 
Deutsch-Jozsa algorithm~\cite{qDJ92}. This requires certain novel combinatorial
observations related to symmetric Boolean functions. Then the quantum state 
out of Deutsch-Jozsa algorithm is measured using the parity measurement 
technique~\cite{d2} to obtain $|D^n_w\rangle$ with a probability 
$\Omega(\frac{1}{\sqrt{n}})$. Thus, $O(\sqrt{n})$ runs are sufficient to obtain
the required Dicke state. Note that a direct approach to construct a 
symmetric state has been presented in~\cite{dickold} using biased Hadamard 
transform. While the order of probability to obtain Dicke state by ours and 
that of~\cite{dickold} are the same, enumeration results show that the exact 
probability values are better in our case than that of~\cite{dickold}.

Further, motivated by the idea in~\cite{dickold}, we improve our algorithm 
further with a modified Deutsch-Jozsa operator that involves the biased 
Hadamard transform. Since biased Hadamard transform also helps to generate the 
target symmetric state, the overall probability to obtain the Dicke state 
increases.

Finally, we can also apply the Grover operator~\cite{qGR96} before the 
measurement. Since Grover algorithm amplifies the amplitude of target 
symmetric state, this helps to reduce the necessary number of steps into 
$O(\sqrt[4]{n})$.

\section{Properties of Symmetric Boolean Functions}
\label{subbool}

\subsection{Walsh Spectrum of Symmetric Boolean Functions}

A Boolean function on $n$ variables may be viewed as a mapping from $\{0,1\}^n$ into $\{0,1\}$. Let us denote the addition operator over $GF(2)$ by $\oplus$. Let ${x}=(x_1,\ldots,x_n)$ and ${\omega}=(\omega_1,\ldots,\omega_n)$ both belong to $\{0,1\}^n$ and the inner product 
$${x}\cdot{\omega}=x_1\omega_1\oplus\cdots\oplus x_n\omega_n.$$ 
Let $f({x})$ be a Boolean function on $n$ variables. Then the {\em Walsh 
transform} of $f({x})$ is a real valued function over $\{0,1\}^n$ which is 
defined as 
$$W_f({\omega})=\sum_{{x}\in\{0,1\}^n}(-1)^{f({x})\oplus{x}\cdot{\omega}}.$$

An $n$-variable Boolean function $f$ is called Symmetric if $f(x)=f(y)$ for all $x, y \in\{0,1\}^n$ such that $wt(x)=wt(y)$. Henceforth, we will denote the set of $n$-variable symmetric Boolean functions as ${\cal SB}_n$. 

In the truth table of $f \in {\cal SB}_n$, it is enough to provide outputs corresponding to different weights of elements of $\{0, 1\}^n$ only. So an $n$-variable symmetric function can be expressed by an $(n+1)$ length bit string as
$$re_f=[f_0,f_1,\ldots,f_n],$$
where $f_i$ is the output at the inputs of weight $i$ and $re_f$ is referred to as the simplified value vector. When $f\in{\cal SB}_n$, one may note that $W_f(x)=W_f(y)$ for all $x,y\in\{0,1\}^n$ such that $wt(x) = wt(y)$. Therefore, the Walsh spectrum of $f$ can be represented by an $(n+1)$ length integer string 
$$rw_f=[rw_f(0),rw_f(1),\ldots,rw_f(n)],$$
where $rw_f(i)$ represents the Walsh spectrum value at the inputs of weight $i$.

\subsection{Relation between Walsh Spectrum and Krawtchouk polynomials}
We now relate the Walsh spectrum of the symmetric functions~\cite{kws} with 
Krawtchouk polynomials~\cite{IK96,FJ77}. Krawtchouk polynomial of degree $i$ is 
given by $$K_i(\eta,n)=\sum_{j=0}^i(-1)^j\binom{\eta}{j}\binom{n-\eta}{i-j}.$$ 
From~\cite{kws}, we get that if $wt(\omega)=k$, then 
$$W_f(\omega)=\sum_{i=0}^n(-1)^{f_i}K_i(k, n).$$ The $(n+1)\times(n+1)$ matrix 
which has $K_i(k,n)$ as the $(i,k)$-th element is known as the Krawtchouk 
matrix~\cite{krawten,fein}. 

For example, let us present the Krawtchouk matrix for $n = 5$ and $6$
as follows:
\begin{center}
$\left[
\begin{array}{rrrrrr}
 1   &   1   &   1   &   1   &   1   &    1 \\
 5   &   3   &   1   &  -1   &  -3   &   -5 \\
10   &   2   &  -2   &  -2   &   2   &   10 \\
10   &  -2   &  -2   &   2   &   2   &  -10 \\
 5   &  -3   &   1   &   1   &  -3   &    5 \\
 1   &  -1   &   1   &  -1   &   1   &   -1 
\end{array}
\right],$
$\left[
\begin{array}{rrrrrrr}
1  &  1  &  1  &  1  &  1  &   1  &  1 \\
6  &  4  &  2  &  0  & -2  &  -4  & -6 \\
15 &  5  & -1  & -3  & -1  &   5  & 15 \\
20 &  0  & -4  &  0  &  4  &   0  & -20 \\
15 & -5  & -1  &  3  & -1  &  -5  & 15 \\
6  & -4  &  2  &  0  & -2  &   4  & -6 \\
1  & -1  &  1  & -1  &  1  &  -1  &  1 
\end{array}
\right].$
\end{center}
In these two matrices, one can verify the properties related to the
Krawtchouk matrix given in Proposition \ref{prop3a}.

To determine all the Walsh spectrum values of $f\in{\cal SB}_n$, it is enough 
to multiply $((-1)^{f_0},\ldots,(-1)^{f_n})$ with the $(n+1)\times(n+1)$ 
Krawtchouk matrix. Applying Krawtchouk matrix, the analysis of the Walsh 
spectra of symmetric functions becomes combinatorially interesting. Elements 
of a Krawtchouk matrix have nice combinatorial properties and they follow nice 
symmetry~\cite{IK96} too. We list some of them in the following proposition.

\begin{proposition}
\label{prop3a}
1. $K_0(k, n)=1, K_1(k, n) = n-2k$,\\
2. $(i+1)K_{i+1}(k, n)=(n-2k)K_i(k,n)$\\
   $-(n-i+1)K_{i-1, n}(k, n)$,\\
3. $K_i(k, n)= (-1)^k K_{n-i}(k, n)$,\\
4. $\binom{n}{k} K_i(k, n)= \binom{n}{i} K_k(i, n)$,\\
5. $K_i(k, n) = (-1)^i K_i(n-k, n)$,\\
6. $(n-k)K_i(k+1, n) = (n-2i) K_i(k, n) - k K_i(k-1, n)$,\\
7. $(n-i+1)K_i(k,n+1)=(3n-2i-2k+1)K_i(k,n)$\\
   $-2(n-k)K_i(k,n-1)$.
\end{proposition}

\subsection{Implementation of Symmetric Boolean Functions}

The symmetric Boolean functions can be efficiently implemented. As described in~\cite{sympoly}, the circuit complexity of $n$-variable symmetric Boolean functions is $4.5n + o(n)$. It is known that given a classical circuit $f$, there is a quantum circuit of comparable efficiency which computes the transformation $U_f$ that takes input like $|x, y\rangle$ and produces output like $|x, y \oplus f(x)\rangle$. Thus, we will consider that for $f \in {\cal SB}_n$, the quantum circuit $U_f$ can be efficiently implemented using $O(n)$ circuit complexity.

\section{Algorithms}

\subsection{Find a Special Symmetric Boolean Function which maximizes the Walsh Spectrum}

Consider that we want to maximize the Walsh spectrum value corresponding to weight $w$ points and naturally, from the property of symmetric functions, all of them will be equal. Now we present an important combinatorial result to show how to find such symmetric Boolean functions.

\begin{theorem} \label{thm2}
Consider $f \in {\cal SB}_n$. The function $f$, represented as $re_f$, for which the Walsh spectrum corresponding to the $w$ weight points will be maximized, can be written as
$$
f_i =     \begin{cases}
                    0,          & $if$\,\, K_i(w, n) > 0 \\
                    1,          & $if$\,\, K_i(w, n) < 0 \\
                    0\, $or$\, 1,   & $if$\,\, K_i(w, n) = 0
            \end{cases}
$$
\end{theorem}
\pf We have $W_f(\omega) = \sum_{i = 0}^n (-1)^{f_i} K_i(k, n)$. One may note 
that the maximum value of $\sum_{i = 0}^n (-1)^{f_i} K_i(k, n)$ is 
$\sum_{i = 0}^n |K_i(k, n)|$. This is attained when we take the function of the form as described in the theorem.
\qed

\begin{example}
\label{ex1}
As example, consider $n = 6, w = k = 2$. In the corresponding column of 
the $(6+1) \times (6+1)$ matrix, we get the values as $1, 2, -1, -4, -1, 2, 1$.
Thus, we will consider the function with $re_f$ as $[0, 0, 1, 1, 1, 0, 0]$.
For such an $f \in {\cal SB}_6$, the Walsh spectrum values at the points 
$\omega$, such that $w = wt(\omega) = 2$, will be maximized, which is
$1+2+1+4+1+2+1 = 12$.
\end{example}

\subsection{Walsh Spectrum of the Special Symmetric Boolean function by Combinatorial Property of Krawtchouk Matrix}

Next we present certain results related to column sum of Krawtchouk matrix.
\begin{lemma}\label{lem1}
$\sum_{i=0}^n |K_i(\lceil\frac{n}{2}\rceil, n)| = \sum_{i=0}^n |K_i(\lfloor\frac{n}{2}\rfloor, n)| =2^{\lceil\frac{n}{2}\rceil}$.
\end{lemma}
\pf Let us first prove this for even $n$.

Following Proposition~\ref{prop3a}(2), we have 
\begin{eqnarray*}
(i+1)K_{i+1}(k, n) & = & (n-2k) K_i(k, n)\\
                   &   & -(n-i+1)K_{i-1, n}(k, n). 
\end{eqnarray*}
For $n$ even, and $k = \frac{n}{2}$, we get, 
$$K_{i+1}(\frac{n}{2}, n) = -\frac{n-i+1}{i+1} K_{i-1, n}(\frac{n}{2}, n).$$ 
That is, the recurrence relation follows: 
$$K_{i}(\frac{n}{2}, n) = -\frac{n-i+2}{i} K_{i-2, n}(\frac{n}{2}, n),$$ with 
the initial conditions $K_0(\frac{n}{2}, n) = 1$ and $K_1(\frac{n}{2}, n) = 0$ 
as available from Proposition~\ref{prop3a}(1). Thus one may note that for odd 
$i$, $K_i(\frac{n}{2}, n) = 0$. Further, using induction, for even $i$, we get 
$$|K_i(\frac{n}{2}, n)| = \binom{\frac{n}{2}}{\frac{i}{2}}.$$ 
Thus, $$\sum_{i=0}^n |K_i(\frac{n}{2}, n)|= 
\sum_{l=0}^{\frac{n}{2}}\binom{\frac{n}{2}}{l},$$ putting $i = 2l$. Hence, 
$$\sum_{i=0}^n |K_i(\frac{n}{2}, n)| = 2^{\frac{n}{2}}.$$
Now let us prove this for odd $n$.

For $n$ odd and $k = \frac{n-1}{2}$, from Proposition~\ref{prop3a}(2) we get 
\begin{eqnarray*}
K_{i+1}(\frac{n-1}{2},n) & = & \frac{1}{i+1}K_i (\frac{n-1}{2},n)\\
 & & - \frac{n-i+1}{i+1}K_{i-1}(\frac{n-1}{2},n). 
\end{eqnarray*}
That is, the recurrence relation is as follows:
\begin{eqnarray*}
K_{i}(\frac{n-1}{2},n) & = & \frac{1}{i}K_{i-1} (\frac{n-1}{2},n)\\
& &  - \frac{n-i+2}{i}K_{i-2}(\frac{n-1}{2},n). 
\end{eqnarray*}
One can now show by induction that 
$$K_{2i}(\frac{n-1}{2},n) = K_{2i+1}(\frac{n-1}{2},n),\ \forall i, \ 1 \leq i \leq \frac{n-1}{2}.$$ Using the above two identities and induction, 
one can verify that $|K_{2l}(\frac{n-1}{2},n)| = \binom{\frac{n-1}{2}}{l}$. 
Thus, $$\sum_{i=0}^n |K_i(\lfloor\frac{n}{2}\rfloor, n)| = 2\sum_{l=0}^{\frac{n-1}{2}} \binom{\frac{n-1}{2}}{l},$$ where $i = 2l$. Hence, we get, 
$$\sum_{i=0}^n |K_i(\lfloor\frac{n}{2}\rfloor, n)| = 2 \cdot 2^{\frac{n-1}{2}} = 2^{\frac{n+1}{2}}.$$

Using Proposition~\ref{prop3a}(5), we get that 
$$\sum_{i=0}^n |K_i(\lfloor\frac{n}{2}\rfloor, n)| = \sum_{i=0}^n |K_i(\lceil\frac{n}{2}\rceil, n)|.$$ That completes the proof. \qed

\begin{theorem}\label{thm3}
Let $f \in {\cal SB}_n$ be as explained in Theorem~\ref{thm2} towards 
maximizing the Walsh spectrum values at weight $\lceil\frac{n}{2}\rceil$ or 
$\lfloor\frac{n}{2}\rfloor$. Then, $$\binom{n}{\lceil\frac{n}{2}\rceil} 
(rw_f(\lceil\frac{n}{2}\rceil))^2 = \binom{n}{\lfloor\frac{n}{2}\rfloor} 
(rw_f(\lfloor\frac{n}{2}\rfloor))^2 \mbox{ is } \Omega(\frac{2^{2n}}{\sqrt{n}}).$$
\end{theorem}
\pf
The Walsh spectrum in this case is 
\begin{eqnarray*}
rw_f(\lceil\frac{n}{2}\rceil) & = & \sum_{i=0}^n |K_i(\lceil\frac{n}{2}\rceil, n)|\\
= rw_f(\lfloor\frac{n}{2}\rfloor) & = & 
\sum_{i=0}^n |K_i(\lfloor\frac{n}{2}\rfloor, n)|. 
\end{eqnarray*}
Thus the total sum of the squares of the Walsh spectrum values at weight 
$\lceil\frac{n}{2}\rceil$ or $\lfloor\frac{n}{2}\rfloor$ is 
$$\binom{n}{\lceil\frac{n}{2}\rceil}(\sum_{i=0}^n |K_i(\lceil\frac{n}{2}\rceil, n)|)^2$$ 
which is $\Omega(\frac{2^{2n}}{\sqrt{n}})$, by Stirling's approximation. \qed

\begin{figure}[t]
\centering
\includegraphics[width=0.49\textwidth]{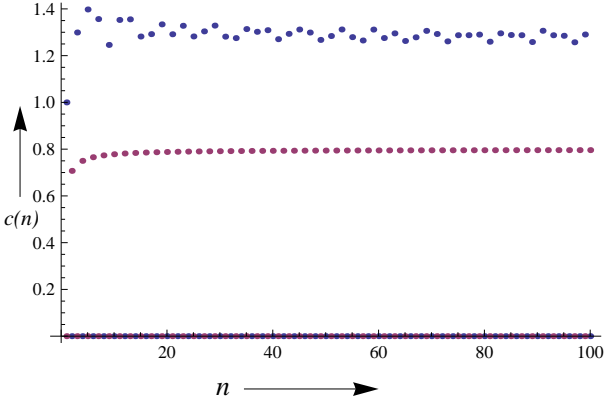}
\caption{Plot of $c(n)$ vs $n$ for $1 \leq n \leq 100$.}
\label{dickplot}
\end{figure}

One may similarly note that for the trivial cases of $w = 0$ or $n$, if one chooses $f \in {\cal SB}_n$ following Theorem~\ref{thm2}, then $\binom{n}{w} (rw_f(w))^2 = 2^{2n}$. However, proving the result similar to Theorem~\ref{thm3} for any $n$ and any weight $w$, in general, seems to be quite tedious. Thus we make detailed enumerations to obtain $c(n) = \min_{w = 0}^n \frac{\binom{n}{w} \left(rw_f(w)\right)^2}{\frac{2^{2n}}{\sqrt{n}}}$ that has been verified for
$n \leq 1000$ and we note that the values stabilize as $c(999) = 1.24793$ and $c(1000) = 0.797685$. The graph of this is plotted in Figure~\ref{dickplot} for $1 \leq n \leq 100$, the points for odd $n$ are coming above and those for even $n$ are coming below. Since we are not providing a proof of this, we refer this 
as follows.

\begin{fact}
\label{fact1}
Let $f \in {\cal SB}_n$ be as explained in Theorem~\ref{thm2} towards maximizing the Walsh spectrum values at weight $w$.
Then the total sum of the squares of the Walsh spectrum values at weight $w$, $\binom{n}{w} \left(rw_f(w)\right)^2$, is $\Omega(\frac{2^{2n}}{\sqrt{n}})$.
\end{fact}
The proof of the fact seems to be quite tedious and elusive and we leave it 
as an open problem.

\subsection{Relation between Deutsch-Jozsa algorithm and the Walsh Spectrum of Symmetric Boolean Function}

Given $f$ is either constant or balanced, if the corresponding quantum implementation $U_f$ is available, Deutsch-Jozsa~\cite{qDJ92} provided a quantum algorithm that decide in constant time which one it is. Let us now describe our interpretation of Deutsch-Jozsa algorithm in terms of Walsh spectrum values. We denote the operator for Deutsch-Jozsa algorithm as 
\begin{equation}
{\cal D}_f = H^{\otimes n} U_f H^{\otimes n}, \nonumber
\end{equation}
where the Boolean function $f$ is available as an oracle $U_f$. For brevity, we abuse the notation and do not write the auxiliary qubit, i.e., $\frac{|0\rangle - |1\rangle}{\sqrt{2}}$ and the corresponding output in this case. 

Now one can observe that~\cite{maitra}
\begin{eqnarray}\nonumber
{\cal D}_f |0\rangle^{\otimes n}    &   =   &   \sum_{z \in \{0, 1\}^n}\sum_{x \in \{0, 1\}^n} \frac{(-1)^{x \cdot z \oplus f(x)}}{2^n}|z\rangle \\ \nonumber
                                    &   =   &   \sum_{z \in \{0, 1\}^n} \frac{W_f(z)}{2^n} |z\rangle. \nonumber
\end{eqnarray}
Note that the associated probability with a state $|z\rangle$ is $\frac{W_f^2(z)}{2^{2n}}$. Hence we have the following technical result as pointed out in~\cite{maitra} with our interpretation for symmetric functions.

\begin{proposition}
\label{prop1}
Given an $n$-variable Boolean function $f$, ${\cal D}_f |0\rangle^{\otimes n}$ produces a superposition of all states $z \in \{0, 1\}^n$ with the amplitude $\frac{W_f(z)}{2^n}$ corresponding to each state $|z\rangle$. Specially, if $f \in {\cal SB}_n$, then the amplitude corresponding to $|z\rangle$ is $\frac{\sum_{i = 0}^n (-1)^{f_i} K_i(wt(z), n)}{2^n}$.
\end{proposition}

\subsection{Algorithm 1: Deutsch-Jozsa Algorithm with Special Symmetric Function}
\label{dj}

Based on the overall properties, we propose a quantum algorithm as shown in the following algorithm.\\

\noindent {\bf Algorithm 1}\\
\begin{tabular}{p{240pt}} \hline
1. Choose $f \in {\cal SB}_n$ as explained in Theorem~\ref{thm2} to maximize the Walsh spectrum values at weight $w$. \\
2. Use the Deutsch-Jozsa algorithm to obtain a symmetric $n$-qubit state $|S^n\rangle = \sum\limits_{x \in \{0, 1\}^n} a_{wt(x)} |x\rangle$, such that $\binom{n}{w} |a_w|^2$ is $\Omega(\frac{1}{\sqrt{n}})$.\\
3. Apply the parity measurement strategy. If the ancilla state is measured at the basis $U^w|\zeta\rangle$, then $|D_w^n\rangle$ is successfully obtained. Else go to Step 2 and iterate.\\   \hline
\end{tabular}\\ \\

The following result provides the estimate of complexity of our algorithm.
\begin{theorem}
\label{thm5}
Let $f \in {\cal SB}_n$ be as explained in Theorem~\ref{thm2} towards maximizing the Walsh spectrum values at weight $w$. Given Fact~\ref{fact1}, Deutsch-Jozsa algorithm produces a symmetric $n$-qubit state (before the measurement) $|S^n\rangle = \sum\limits_{x \in \{0, 1\}^n} a_{wt(x)} |x\rangle$, such that $\binom{n}{w} |a_w|^2$ is $\Omega(\frac{1}{\sqrt{n}})$.
\end{theorem}
\pf
The proof follows from Theorem~\ref{thm2}, Fact~\ref{fact1} and Proposition~\ref{prop1}.
\qed

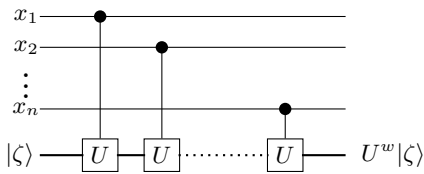
\begin{figure}[t]
\begin{center}
\begin{tikzpicture}[scale=0.82]
\node [inner sep=0pt,outer sep=0pt](p1) at (2,2.5) {};
\node [inner sep=0pt,outer sep=0pt](p2) at (7,2.5) {};
\draw [-] (p1) -- (p2);
\node [inner sep=0pt,outer sep=0pt](p1) at (2,3.5) {};
\node [inner sep=0pt,outer sep=0pt](p2) at (7,3.5) {};
\draw [-] (p1) -- (p2);
\node [inner sep=0pt,outer sep=0pt](p1) at (2,4) {};
\node [inner sep=0pt,outer sep=0pt](p2) at (7,4) {};
\draw [-] (p1) -- (p2);
\node [inner sep=0pt,outer sep=0pt](p1) at (3,4) {};
\node [inner sep=0pt,outer sep=0pt](p2) at (3,2) {};
\draw [-] (p1) -- (p2);
\node [inner sep=0pt,outer sep=0pt](p1) at (4,3.5) {};
\node [inner sep=0pt,outer sep=0pt](p2) at (4,2) {};
\draw [-] (p1) -- (p2);
\node [inner sep=0pt,outer sep=0pt](p1) at (6,2.5) {};
\node [inner sep=0pt,outer sep=0pt](p2) at (6,2) {};
\draw [-] (p1) -- (p2);
\node [inner sep=0pt,outer sep=0pt](p1) at (2,1.75) {};
\node [inner sep=0pt,outer sep=0pt](p2) at (2.75,1.75) {};
\draw [thick] (p1) -- (p2);
\node [inner sep=0pt,outer sep=0pt](p1) at (3.25,1.75) {};
\node [inner sep=0pt,outer sep=0pt](p2) at (3.75,1.75) {};
\draw [thick] (p1) -- (p2);
\node [inner sep=0pt,outer sep=0pt](p1) at (4.25,1.75) {};
\node [inner sep=0pt,outer sep=0pt](p2) at (5.75,1.75) {};
\draw [thick,dotted] (p1) -- (p2);
\node [inner sep=0pt,outer sep=0pt](p1) at (6.25,1.75) {};
\node [inner sep=0pt,outer sep=0pt](p2) at (7,1.75) {};
\draw [thick] (p1) -- (p2);
\node[rectangle, draw=black, minimum height=2, minimum width=2] (H) at (3,1.75) {$U$};
\node[rectangle, draw=black, minimum height=2, minimum width=2] (H) at (4,1.75) {$U$};
\node[rectangle, draw=black, minimum height=2, minimum width=2] (H) at (6,1.75) {$U$};
\fill[black] (3, 4) circle (0.1cm);
\fill[black] (4, 3.5) circle (0.1cm);
\fill[black] (6, 2.5) circle (0.1cm);
\node [inner sep=0pt,outer sep=0pt] (t2) at (1.8,4) {$x_1$};
\node [inner sep=0pt,outer sep=0pt] (t2) at (1.8,3.5) {$x_2$};
\node [inner sep=0pt,outer sep=0pt] (t2) at (1.8,3) {\Large $\vdots$};
\node [inner sep=0pt,outer sep=0pt] (t2) at (1.8,2.5) {$x_n$};
\node [inner sep=0pt,outer sep=0pt] (t2) at (1.7,1.75) {$|\zeta\rangle$};
\node [inner sep=0pt,outer sep=0pt] (t2) at (7.75,1.75) {$U^w|\zeta\rangle$};
\end{tikzpicture}
\caption{Generalized Parity Module as in~\cite[Fig. 1]{d2}. If $wt(x) = w$, then the ancilla will become $U^w|\zeta\rangle$.}
\label{fig1}
\end{center}
\end{figure}

Now the final step is to measure the symmetric state until we get the target Dicke state by using parity measurement method ~\cite[Section IIIA]{d2}. Now we explain how to exploit the parity measurement in our purpose. Note \cite[Section IIIA]{d2} assumes to use $n$ dimensional qudit ancilla, but we consider a qudit $|\zeta\rangle$ of dimension $n+1$ here. A certain unitary operator $U$ is designed such that, $$|\zeta\rangle, U|\zeta\rangle, U^2|\zeta\rangle, \ldots, 
U^{n-1}|\zeta\rangle, U^{n}|\zeta\rangle$$ are all orthogonal to each other 
and $U^{n+1}|\zeta\rangle = |\zeta\rangle$. 
Since $|\zeta\rangle$ is an $(n+1)$-dimensional state, one can indeed obtain a 
set of such $n+1$ orthogonal states. The parity measurement is done on the 
$$\{|\zeta\rangle, U|\zeta\rangle, U^2|\zeta\rangle, \ldots, U^{n}|\zeta\rangle\}$$ basis. Here $|\zeta\rangle$ is used as the target state. For the $n$-qubit control state $|x\rangle$, if it has weight $w$ then its corresponding target state, after application of this circuit, will become $U^{w}|\zeta\rangle$ (see Figure~\ref{fig1}). Now consider an $n$-qubit symmetric state as the control input, which is $|\tau\rangle =  \sum\limits_{x \in \{0, 1\}^n} a_w |x\rangle,$ where $w = wt(x)$, $\sum_{i = 0}^n \binom{n}{i} |a_i|^2 = 1$. After applying this circuit, one obtains $\sum\limits_{x \in \{0, 1\}^n} a_w |x\rangle U^w|\zeta\rangle$. Thus, if one measures in $$\{|\zeta\rangle, U|\zeta\rangle, U^2|\zeta\rangle, \ldots, U^{n}|\zeta\rangle\}$$ basis, then the state $|D_w^n\rangle$ will be obtained when the measurement result of the ancilla state is $U^w|\zeta\rangle$.

Since the probability of target Dicke state is $\Omega(\frac{1}{\sqrt{n}})$, we should repeat the whole procedure at most $O(\sqrt{n})$.

\begin{example}
\label{ex2}
Let us have an example taking $n = 6, w = 2$ to outline our method.
In this case the Dicke state will be
$$|D^6_2\rangle = \sum\limits_{x \in \{0, 1\}^6, wt(x) = 2} 
\frac{1}{\sqrt{15}} |x\rangle.$$ We start with is an $n = 6$ variable symmetric 
Boolean function having Walsh spectrum value at each of the weight $w = 2$
point as $12$ (following Theorem~\ref{thm2}, one may refer to Example~\ref{ex1}
also). There are
$\binom{n}{w} = \binom{6}{2} = 15$ such points. The amplitude associated with
each point after the DJ algorithm is $\frac{12}{2^6} = \frac{3}{16}$.
Thus we get $\sum_{x \in \{0, 1\}^6, wt(x) = 2} 
\frac{3}{16} |x\rangle + \sum_{x \in \{0, 1\}^6, wt(x) \neq 2} 
a_x |x\rangle$ initially. Thus, the probability associated with $|D^6_2\rangle$
will be $\binom{6}{2} (\frac{3}{16})^2 = \frac{135}{256} = 0.52734375$ and 
hence one may note that the probability that, after parity measurement, it 
will land into $|D^6_2\rangle$ is quite high.
\end{example}

\subsection{Comparison with a Previous Method}
It was explained in~\cite{d2} how one can obtain $|D^n_w\rangle$ from 
$\frac{1}{2^{\frac{n}{2}}} \sum_{x \in \{0, 1\}^n} |x\rangle$. However, the 
idea explained in~\cite[Section IIIA]{d2} works efficiently only for 
$w = \lceil\frac{n}{2}\rceil$ or $\lfloor\frac{n}{2}\rfloor$. The most general 
work in this direction has appeared in~\cite{dickold} where biased Hadamard
transformation was exploited.
The strategy of~\cite{dickold} uses biased Hadamard transformation
$$\begin{pmatrix}
\sqrt{1-\frac{w}{n}} & \sqrt{\frac{w}{n}} \\
\sqrt{\frac{w}{n}} & -\sqrt{1-\frac{w}{n}}
\end{pmatrix}^{\otimes n}$$ on $|0\rangle^{\otimes n}$ such that the probability associated with $|D^n_w\rangle$ will be
$\binom{n}{w} (\frac{w}{n})^w (1-\frac{w}{n})^{n-w} \geq \sqrt{\frac{2}{n\pi}}$, i.e., $\Omega(\frac{1}{\sqrt{n}})$. Thus, the probability of our case and also in~\cite{dickold} are of the same order. While the theoretical comparison of the exact probability values seems elusive, we have made detailed enumerations to observe that the exact probability values in our case are better than that 
of~\cite{dickold}. Below we present enumeration results towards that. 

\begin{figure}[t]
\centering
\includegraphics[width=0.49\textwidth]{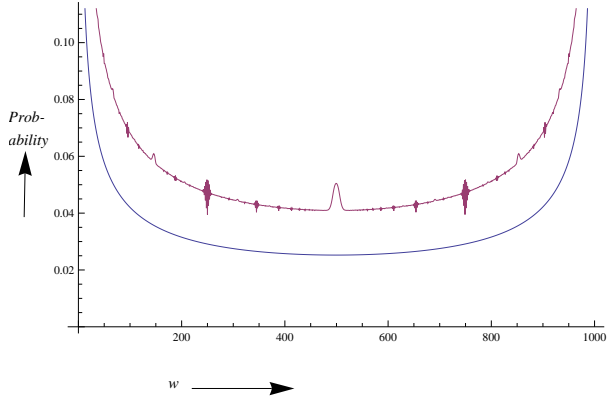}
\caption{Plot of probabilities associated with $|D^n_w\rangle$ against $w$ in our case (above) and in~\cite{dickold} (below) for $n = 999$.}
\label{comp1a}
\end{figure}

\begin{figure}[t]
\centering
\includegraphics[width=0.49\textwidth]{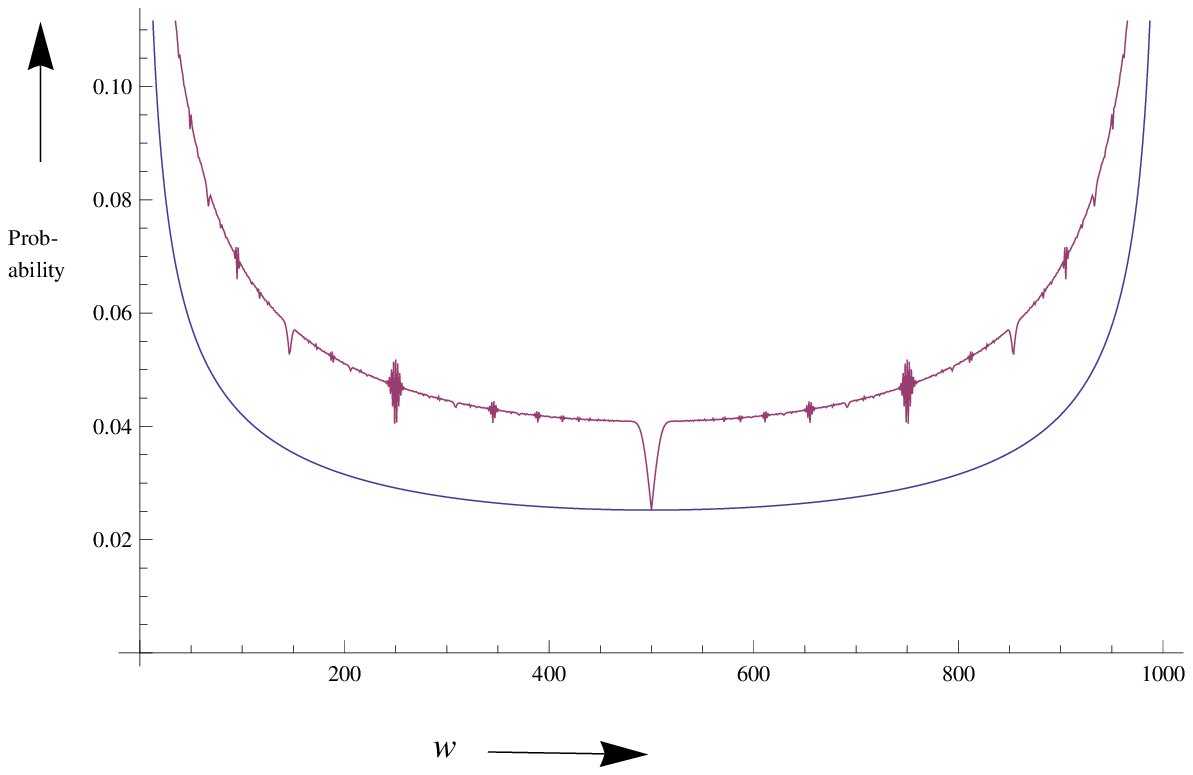}
\caption{Plot of probabilities associated with $|D^n_w\rangle$ against $w$ in our case (above) and in~\cite{dickold} (below) for $n = 1000$.}
\label{comp1}
\end{figure}

First we present two graphs to show the probability values associated to $|D^n_w\rangle$ for all $w$, when $n = 999$ (to represent odd case) or $1000$ (to represent even case). For our case, it is $\binom{n}{w} \frac{(\sum_{i = 0}^n |K_i(w, n)|)^2}{2^{2n}}$ (after application of Deutsch-Jozsa algorithm without measurement), and for the case of \cite{dickold} it is $\binom{n}{w} (\frac{w}{n})^w (1-\frac{w}{n})^{n-w}$ (after application of biased Hadamard transform). From Figure~\ref{comp1a} and Figure~\ref{comp1}, it is clear to note that our method provides higher probability (the upper curve) in all the cases except $w = 0, n$ (which are trivial ones) and $w = \frac{n}{2}$ for $n = 1000$.

\begin{figure}[t]
\centering
\includegraphics[width=0.49\textwidth]{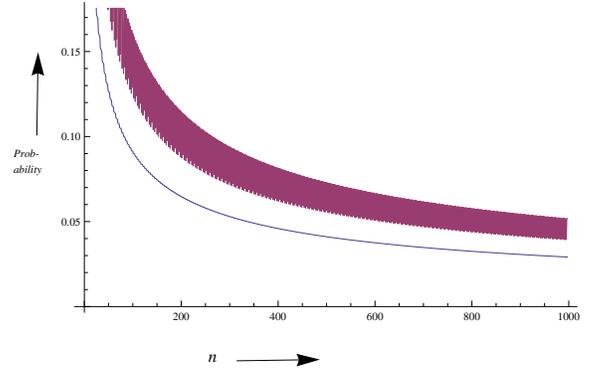}
\caption{Plot of probabilities associated with $|D^n_w\rangle$ against
$n$ in our case (above) and in~\cite{dickold} (below) for $n = 4$ to $1000$
and $w = \lfloor \frac{n}{4}\rfloor$.}
\label{comp2}
\end{figure}

In both figures, the present algorithm shows some variation of probability 
when the weight is around $\lfloor \frac{n}{4} \rfloor$. To check whether or 
not these cases still shows the higher probability than the previous method, 
we look into the $w = \lfloor \frac{n}{4} \rfloor$ case a little bit more. 
From Figure~\ref{comp2}, one may note that our probability values (the upper curve) are higher than the case of~\cite{dickold}. These results explains the advantage of the use of a suitable symmetric Boolean function which shows higher Walsh spectrum values for the given weight.

\subsection{Algorithm 2: Additional Improvement by exploiting biased Hadamard Operator}

We have provided numerical evidences that using proper symmetric Boolean functions in Deutsch-Jozsa algorithm provides better probability than the use of biased Hadamard transform as described in~\cite{dickold}. However, motivated 
by~\cite{dickold}, a natural extension should be to couple biased Hadamard transform in Deutsch-Jozsa algorithm instead of (unbiased) Hadamard transform. Thus, let us refer to the general description of a Hadamard type transformation (biased or unbiased) that can be written as
$B_{r, n} =
\begin{pmatrix}
\sqrt{1-\frac{r}{n}} & \sqrt{\frac{r}{n}} \\
\sqrt{\frac{r}{n}} & -\sqrt{1-\frac{r}{n}}
\end{pmatrix}$.
We will replace the standard notation of $w$ here by $r$ as we will not restrict ourselves to integer values $w \in [0, \ldots, n]$, but use any real number $r \in [0, n]$ to obtain the optimum probability of success to get a Dicke state.

Instead of using the operator ${\cal D}_f = H^{\otimes n} U_f H^{\otimes n}$, let us first describe the most general operator of the form 
\begin{equation}
{\cal D'}_f = B_{r_1, n}^{\otimes n} U_f B_{r_2, n}^{\otimes n},
\end{equation}
where $r_1, r_2$ are real numbers in $[0, n]$.

First, we consider the case when $r_1 = \frac{n}{2}$, i.e., $B_{r_1, n} = H$, but $r_2 = r$ varies towards optimization. That is 
\begin{equation}
{\cal D'}_f = H^{\otimes n} U_f B_{r, n}^{\otimes n}.
\end{equation}

One may note that the application of ${\cal D'}_f$ on $|0\rangle^{\otimes n}$ will produce

${\cal D'}_f |0\rangle^{\otimes n} = \frac{1}{\sqrt{2^n}} \sum_{z \in \{0, 1\}^n}$

$\left(\sum_{x \in \{0, 1\}^n} (-1)^{x\cdot z\oplus f(x)} (1-\frac{r}{n})^{\frac{n-d(x,z)}{2}} (\frac{r}{n})^{\frac{d(x,z)}{2}}\right) |z\rangle$, where $d(x, z)$ is the (Hamming) distance between two same length binary strings $x$ and $z$.
Before proceeding further, we also have the following technical result.
\begin{proposition}
Let ${\cal D'}_f = H^{\otimes n} U_f B_{r, n}^{\otimes n}$.
If $f \in {\cal SB}_n$ then ${\cal D'}_f |0\rangle^{\otimes n}$
is a symmetric state.
\end{proposition}
\pf We need to prove that
$\sum_{x \in \{0, 1\}^n} (-1)^{x\cdot z\oplus f(x)} 
(1-\frac{r}{n})^{\frac{n-d(x,z)}{2}} (\frac{r}{n})^{\frac{d(x,z)}{2}}$
is same for all the $z \in \{0, 1\}^n$ having the same Hamming weight.
Let us consider $u, v \in \{0, 1\}^n$ such that $u \neq v$, but 
$wt(u) = wt(v)$. Then we need to prove that 
$\sum_{x \in \{0, 1\}^n} (-1)^{x\cdot u\oplus f(x)} 
(1-\frac{r}{n})^{\frac{n-d(x,u)}{2}} (\frac{r}{n})^{\frac{d(x,u)}{2}}$
$=$
$\sum_{x \in \{0, 1\}^n} (-1)^{x\cdot v\oplus f(x)} 
(1-\frac{r}{n})^{\frac{n-d(x,v)}{2}} (\frac{r}{n})^{\frac{d(x,v)}{2}}$.

The proof follows from the fact that 
$\sum_{x \in \{0, 1\}^n, wt(x) = w} (-1)^{x\cdot u\oplus f(x)} 
(1-\frac{r}{n})^{\frac{n-d(x,u)}{2}} (\frac{r}{n})^{\frac{d(x,u)}{2}} =$
$\sum_{x \in \{0, 1\}^n, wt(x) = w} (-1)^{x\cdot v\oplus f(x)} 
(1-\frac{r}{n})^{\frac{n-d(x,v)}{2}} (\frac{r}{n})^{\frac{d(x,v)}{2}}$,
given that $f$ is symmetric. \qed

The main problem in this case is that we need to go for trial and error by
modifying the symmetric Boolean functions and trying out different values 
of $\frac{r}{n}$. So far, we could not obtain the exact characterization 
of symmetric functions while biased Hadamard transform is used.
Based on this, we propose Algorithm 2 as follows.\\

\noindent {\bf Algorithm 2}\\
\begin{tabular}{p{240pt}} \hline
1. Apply ${\cal D'}_f$ to $|0\rangle^{\otimes n}$ to obtain a symmetric $n$-qubit state. The value of $\frac{r}{n}$ and the choice of the symmetric Boolean 
functions are achieved heuristically.\\
2. Use parity measurement strategy. If the ancilla state is measured at the basis $U^w|\zeta\rangle$, then $|D_w^n\rangle$ is successfully obtained. Else take
the parameters as in Step 1 once again and iterate Step 2.\\   \hline
\end{tabular}\\ \\

As we could not characterize this, to provide some experimental results in this
direction (see Table \ref{comparison_three} at the end of this draft), 
we used the 
following method
for some small values of $n$ ($n = 4$ to $9$). We select each of the Boolean 
functions $f$ from ${\cal SB}_n$. Given $f$ and a specific weight $w$, 
$1\leq w\leq n$, we write the expression of success probability as a function 
of $r$. Then we apply the $Maximize$ function available in $Mathematica$ $8.0$ 
to compute the optimum value of $r$ given $f, w$, so that the success 
probability becomes maximum. Note that, as we could not characterize the 
process yet, this is an exhaustive task and for each $n$, it requires checking 
of $2^{n+1}$ symmetric Boolean functions.
That is the reason, we can study it for only a few small values of $n$. However,
this is a classical computation that can be done as an off-line work. Once such
programs are executed, we can have a database of proper ${f \in \cal SB}_n$
and the corresponding $r$ to have the optimal success probability to obtain
$|D^n_w\rangle$. Given these data, the actual quantum algorithm to obtain
Dicke states can be efficiently implemented.

\section{Numerical Comparison of Three Approaches}
Now we compare three approaches: 
\begin{itemize}
\item using biased Hadamard operator as shown in \cite{dickold}, 
\item Algorithm 1 based on ${\cal D}_f |0\rangle^{\otimes n}$, and 
\item  Algorithm 2 based on ${\cal D'}_f |0\rangle^{\otimes n}$. 
\end{itemize}
The first two cases need $O(\sqrt{n})$ complexity, and the third one is a 
heuristic that shows improved results than the first two.
Some numerical results of the probability associated with 
$|D^n_w\rangle$ are shown in Table \ref{comparison_three} at the end of this
draft for
$n =4, \cdots, 9$. As shown in the table, we note that Algorithm 2 provides
the highest probability than others. There are a few interesting observations
from the enumeration results.
\begin{itemize}
\item We note that the improvements using ${\cal D'}_f$ are highly significant 
at $w = \lfloor\frac{n}{2}\rfloor$ or $\lceil\frac{n}{2}\rceil$ and the
significance reduces as $w$ moves away from the middle, i.e., 
towards $w = 1$ or $n-1$.
\item In case of using ${\cal D'}_f$, the success probabilities at $w$ and 
$n-w$ weights are same for all the values of $w$, i.e., $w \geq 1$. However, 
the values of $r$ in those cases are same at $w$ and $n-w$ weights for 
$w \geq 2$ only. 
\end{itemize}

\section{The complete strategy using Grover Algorithm}
\label{grover}
Quadratic improvement by Grover's algorithm~\cite{qGR96} is achieved in 
several applications. We point out here how that can be exploited in
our algorithm. Although we can construct the target Dicke state by measuring 
the intermediate 
quantum state, we may increase the efficiency further by using the amplitude 
amplification method. Based on this, an adiabatic evolution has been used 
towards amplitude amplification of the desired states in~\cite{dickold}, but 
no complexity analysis was shown. In this work, instead, we apply the 
conventional Grover algorithm~\cite{qGR96} as it provides a quadratic speed-up.

Instead of equal superposition 
$|\psi\rangle = H^{\otimes n} |0\rangle^{\otimes n} = \frac{1}{2^{\frac{n}{2}}} \sum_{x \in \{0, 1\}^n} |x\rangle$ in Grover algorithm, we will use the 
symmetric state of the form $|\Psi\rangle = {\cal D}_f (|0\rangle^{\otimes n}) 
= \sum_{x \in \{0, 1\}^n} \frac{W_f(x)}{2^n} |x\rangle$ exploiting the
properly chosen Boolean function $f(x)$, as explained in the previous sections.

Further, towards inverting the phase, we will use another symmetric Boolean 
function $g(x)$, different from $f(x)$, where $g(x) = 1$, when $wt(x) = w$, 
and $g(x) = 0$, otherwise. Based on $g(x)$, we implement the inversion 
operator as ${\cal O}_g$, that inverts the phase of the states $|x\rangle$ 
where $\{ x \in \{0, 1\}^n \,| wt(x) = w\}$. Thus, we consider the operator 
$$G_t = \left[(2 |\Psi \rangle \langle \Psi| - I) {\cal O}_g \right]^t$$ on 
$|\Psi\rangle$ to get $|\Psi_t\rangle$. 

Consider the $n$-qubit state 
$|\Psi\rangle = \sum_{s \in S} u_s |s\rangle + \sum_{s \in \{0, 1\}^n 
\setminus S} v_s |s\rangle$, where $u_s, v_s$ are real and 
$\sum_{s \in S} u^2_s + \sum_{s \in \{0, 1\}^n \setminus S} v^2_s = 1$. 
For brevity, let us represent 
$|\Psi\rangle = \sum\limits_{s \in S} u_s |s\rangle + 
\sum\limits_{s \in \{0, 1\}^n \setminus S} v_s |s\rangle = 
u|X\rangle + v|Y\rangle$. That is, $u^2 = \sum\limits_{s \in S} u^2_s$ 
and $v^2 = \sum\limits_{s \in \{0, 1\}^n \setminus S} v^2_s$.

Let $u = \sin \theta$, $v = \cos \theta$. It is easy to check that
the application of $[(2|\Psi\rangle\langle\Psi| - I){\cal O}_g]^{t}$ operator 
on $|\Psi\rangle$ produces $|\Psi_t\rangle$, in which the probability 
amplitude of $|X\rangle$ is $\sin (2t+1)\theta$.

We will now use such states $|\Psi_t\rangle$ in parity measurement. 
Consider that after the Deutsch-Jozsa algorithm we obtain a symmetric $n$-qubit
state (before the measurement) 
$$|S^n\rangle = \sum\limits_{x \in \{0, 1\}^n} a_{wt(x)} |x\rangle,$$ 
such that $\binom{n}{w} |a_w|^2 = \frac{c}{\sqrt{n}}$, for some constant $c$. 
Thus, we have the initial amplitude of target states, 
$\{ x \in \{0, 1\}^n\, | wt(x) = w\}$, is 
$\sin\theta = \binom{n}{w} |a_w| = \sqrt{\frac{c}{\sqrt{n}}}$. For large $n$, 
one can approximate it as $\theta = \frac{\sqrt{c}}{\sqrt[4]{n}}$ and hence 
we need $t$ iterations of Grover like strategy such that 
$(2t + 1)\theta \approx \frac{\pi}{2}$, i.e., 
$t \approx \frac{\pi\sqrt[4]{n}}{2\sqrt{c}}$. 

Here we have good (almost exact) estimate of $t$, which is not known priori 
for application in 
search algorithms. After the application of Grover like strategy, we will get 
another symmetric $n$-qubit state 
$|T^n\rangle = \sum\limits_{x \in \{0, 1\}^n} a'_{wt(x)} |x\rangle$ such that 
$\binom{n}{w} |a'_w|^2$ will be very close to $1$ and the parity measurement 
will produce $|D^n_w\rangle$ mostly in one step with very high probability. 
Thus the exact strategy is similar to Algorithm 1 (Algorithm 2 can be modified 
with a similar way) in the previous section, where we add one more step as 
follows.\\

\noindent {\bf Algorithm 3}\\
\begin{tabular}{p{240pt}} \hline
1. Let $f \in {\cal SB}_n$ be as explained in Theorem~\ref{thm2} to maximize the Walsh spectrum values at weight $w$. \\
2. Use any of the above three strategies (our strategies exploiting Hadamard 
or biased Hadamard transform or the strategy of~\cite{dickold}) to obtain a 
symmetric $n$-qubit state  
$|S^n\rangle = \sum\limits_{x \in \{0, 1\}^n} a_{wt(x)} |x\rangle$, 
such that $\binom{n}{w} |a_w|^2$ is $\Omega(\frac{1}{\sqrt{n}})$.\\
3. Use $G_t$ on $|S^n\rangle$, $t$ many times, where $t$ is $O(\sqrt[4]{n})$ to obtain $|T^n\rangle = \sum\limits_{x \in \{0, 1\}^n} a'_{wt(x)} |x\rangle$ such that $\binom{n}{w} |a'_w|^2$ is very close to $1$. \\
4. Use parity measurement strategy. If the ancilla state is measured at the basis $U^w|\zeta\rangle$, then $|D_w^n\rangle$ is successfully obtained. Else go to Step 2 and iterate.\\   \hline
\end{tabular}\\ \\

We need $O(\sqrt[4]{n})$ steps using Grover algorithm in each run and 
then a parity measurement should provide $|D_w^n\rangle$. Thus, we get a 
quadratic speed-up (which is quite natural) over just using Deutsch-Jozsa 
algorithm. The number of parity measurement was $O(\sqrt{n})$ in the 
earlier case, once in each iteration. Here it is only a very few (may be 1 in 
most of the cases). 

\section{Conclusion and Open Problems}
\label{conclusion}
In this work, we study several quantum algorithms to construct arbitrary 
Dicke state in a disciplined manner. The key idea is to find a suitable 
symmetric Boolean function for Deutsch-Jozsa algorithm for the given $n$ and 
$w$, use of the Grover algorithm and the generalized parity measurement 
strategy. Further, we show that it is possible to obtain improved results 
using biased Hadamard transform suitably. Our results improve the probabilities
obtained in~\cite{dickold} and thus provide faster method to construct Dicke 
states. The problem open in this area is to characterize the enumeration 
results in case of modifying the Deutsch-Jozsa algorithm with biased Hadamard 
transform. Obtaining the exact bias ($\frac{r}{n}$) in biased Hadamard 
transform with the corresponding symmetric function to optimize the probability
corresponding to the Dicke state seems to be an interesting problem.

Though we look at the problem from theoretical angle, the algorithmic blocks
used by us have experienced major advancement towards actual implementation.
One may refer to~\cite[Section 7]{qNC02} for literature related to 
implementation of quantum gates as well as Deutsch-Jozsa algorithm, Grover 
algorithm and several measurement techniques. As example, the idea of 
implementing biased Hadamard transform is related to the Fabry-Perot 
cavity~\cite[Page 299]{qNC02}.

\ \\
{\bf Acknowledgments:} The authors like to thank Prof. A. M. Childs for 
providing critical comments on an initial version of this paper and pointing
to~\cite{dickold}.

\begin{table*}[t]
\begin{tabular}{|c|c|l|l|l|l|l|l|l|l|} \hline\hline
$n \downarrow$ & $w \rightarrow$  &1 & 2 & 3 & 4 & 5 & 6 & 7 & 8\\
\hline \hline
4 & ${\cal D'}_f |0\rangle^{\otimes n}$ & 0.833609 & 0.981763 & 0.833609 & -- & -- & -- & -- & --\\
  & $f$ & 01 & 02 & 05 & -- & -- & -- & -- & --\\
  & $r$ & 0.468136 & 0.298698 & 0.468136 & -- & -- & -- & -- & --\\ \cline{2-10}
  & ${\cal D}_f |0\rangle^{\otimes n}$ & 0.5625 & 0.375 & 0.5625 & -- & -- & -- & -- & --\\ \cline{2-10}
  & \cite{dickold} & 0.421875 & 0.375 & 0.421875 & -- & -- & -- & -- & --\\ \hline\hline
5 & ${\cal D'}_f |0\rangle^{\otimes n}$ & 0.748304 & 0.92852 &  0.92852 & 0.748304 & -- & -- & -- & --\\
  & $f$ & 03 & 02 & 05 & 16 & -- & -- & -- & --\\
  & $r$ & 1.42458 & 0.313077 & 0.313077 & 3.57542 & -- & -- & -- & --\\ \cline{2-10}
  & ${\cal D}_f |0\rangle^{\otimes n}$ & 0.703125 & 0.625 & 0.625 & 0.703125 & -- & -- & -- & --\\ \cline{2-10}
  & \cite{dickold} & 0.4096 & 0.3456 & 0.3456 & 0.4096 & -- & -- & -- & --\\ \hline\hline
6 & ${\cal D'}_f |0\rangle^{\otimes n}$ & 0.730278 & 0.823495 & 0.954987 & 0.823495 & 0.730278 & -- &-- & --\\
  & $f$ & 03 & 02 & 05 & 0A & 29 & -- & -- & --\\
  & $r$ & 1.48129 & 0.357282 & 0.277975 & 0.357282 & 4.51871 & -- & -- & --\\ \cline{2-10}
  & ${\cal D}_f |0\rangle^{\otimes n}$ & 0.585938 & 0.527344 & 0.3125 & 0.527344 & 0.585938 & -- & -- & --\\ \cline{2-10}
  & \cite{dickold} & 0.401878 & 0.329218 & 0.3125 & 0.329218 & 0.401878 & -- & -- & -- \\ \hline\hline
7 & ${\cal D'}_f |0\rangle^{\otimes n}$ & 0.704306 & 0.754753 & 0.907588 & 0.907588 & 0.754753 & 0.704306 & -- & --\\
  & $f$ & 07 & 60 & 05 & 0A & 53 & 4A & -- & --\\
  & $r$ & 2.44507 & 5.93733 & 0.27984 & 0.27984 & 5.93733 & 2.44507 & -- & --\\ \cline{2-10}
  & ${\cal D}_f |0\rangle^{\otimes n}$ & 0.683594 & 0.512695 & 0.546875 & 0.546875 & 0.512695 & 0.683594 & -- & --\\ \cline{2-10}
  & \cite{dickold} & 0.396569 & 0.318745 & 0.293755 & 0.293755 & 0.318745 & 0.396569 & -- & --\\ \hline \hline
8 & ${\cal D'}_f |0\rangle^{\otimes n}$ & 0.698181 & 0.710643 & 0.813922 & 0.92625 & 0.813922 & 0.710643 & 0.698181 & --\\
  & $f$ & 3F & C0 & BF & A0 & AF & AC & AD & --\\
  & $r$ & 5.51859 & 6.91248 & 7.69903 & 7.74472 & 7.69903 & 6.91248 & 5.51859 & --\\ \cline{2-10}
  & ${\cal D}_f |0\rangle^{\otimes n}$ & 0.598145 & 0.553711 & 0.413574 & 0.273438 & 0.413574 & 0.553711 & 0.598145 & -- \\ \cline{2-10}
  & \cite{dickold} & 0.392696 & 0.311462 & 0.281632 & 0.273438 & 0.281632 & 0.311462 & 0.392696 & --\\ \hline\hline
9 & ${\cal D'}_f |0\rangle^{\otimes n}$ & 0.684842 & 0.651002 & 0.76886 & 0.884277 & 0.884277 & 0.76886 & 0.651002 & 0.684842\\
  & $f$ & 0F & 180 & 0D & 140 & 15F & 6A & 153 & 16A\\
  & $r$ & 3.4566 & 7.86171 & 0.858163 & 8.7469 & 8.7469 & 0.858153 & 7.86171 & 3.4566\\ \cline{2-10}
  & ${\cal D}_f |0\rangle^{\otimes n}$ & 0.672913 & 0.430664 & 0.415283 & 0.492188 & 0.492188 & 0.415283 & 0.430664 & 0.672913\\ \cline{2-10}
  & \cite{dickold} & 0.389744 & 0.306102 & 0.273129 & 0.260182 & 0.260182 & 0.273129 & 0.306102 & 0.389744\\ \hline\hline
\end{tabular}
\caption{Probability values using biased Hadamard transform (in this case we 
provide the corresponding symmetric function $f$ represented as a hexadecimal 
number of the $(n+1)$ length bit-string $f_n,f_{n-1},\ldots,f_1,f_0$ and the 
value of $r$), using (standard) Hadamard transform and the method 
of~\cite{dickold}.}
\label{comparison_three}
\end{table*}

\end{document}